\documentclass[pre,aps,amsmath,eqsecnum,unsortedaddress,nofootinbib,preprint,floatfix]{revtex4-1}

\usepackage{amssymb}
\usepackage{amsmath}
\usepackage{graphicx}
\usepackage{amsbsy}
\usepackage{color}

\begin{document}
\title{Coil-globule transition of a homopolymer chain in a square-well potential: Comparison between Monte Carlo  canonical replica exchange and Wang-Landau sampling.}

\author{Artem Badasyan}
\email{abadasyan@gmail.com}
\affiliation{Department of Theoretical Physics, Jo\v zef Stefan Institute, SI-1000 Ljubljana, Slovenia, EU}

\author{Trinh Xuan Hoang}
\email{hoang@iop.vast.ac.vn}
\affiliation{Center for Computational Physics
Institute of Physics, VAST 10 Dao Tan St., Hanoi, Vietnam}

\author{Rudolf Podgornik}
\email{rudolf.podgornik@ijs.si}
\affiliation{Department of Theoretical Physics, 
J. Stefan Institute and Department of Physics, 
Faculty of Mathematics and Physics, 
University of Ljubljana - SI-1000 Ljubljana, Slovenia, EU}

\author{Achille Giacometti}
\email{achille@unive.it}
\affiliation{Dipartimento di Scienze Molecolari e Nanosistemi, Facolta' di Science, Universita' Ca' Foscari Venezia, I-30123 Venezia, Italy, EU}

\date{\today}

\begin{abstract}
We study the equilibrium properties of a flexible homopolymer where consecutive monomers are represented by impenetrable hard spheres that are tangent to
each other, and non-consecutive monomers interact via a square-well potential. To this aim, we use both replica exchange canonical simulations
and micro-canonical Wang-Landau techniques for relatively short chains, and perform a close comparative analysis of the corresponding results.
These investigations are then further exploited to reproduce, at a much shorter scale and, hence, computational effort, the phase diagram previously
studied with much longer chains. This opens up the possibility of improving the model and introduce specificities typical, among other examples, of protein folding.
\end{abstract}


\maketitle
\section{Introduction}
\label{sec:introduction}
Square-well (SW) potential has a long and venerable tradition in simple liquids \cite{Hansen86}, and has become a paradigmatic test-bench for more sophisticate 
new approaches. Initially it was used as a minimal model, alternative to Lennard-Jones potential, in early attempts of molecular dynamics simulations of liquids 
\cite{Adler59}, because it could be more easily implemented in a simulation code, and yet contained the most important features of pair potential for a liquid.
Indeed, it displays both a gas-liquid and liquid-solid transitions in the phase diagram, with results often quantitatively in agreement with real atomistic fluids
\cite{Barker76}. For sufficiently short-range attraction, the gas-liquid transition becomes metastable and gets pre-empted by a direct gas-solid 
transition \cite{Pagan05,Liu05}. Several variants of the SW model have also been proposed over the years in the framework of molecular fluids \cite{Gray84} and colloidal suspensions \cite{Lyklema91}. 

In the framework of polymer theory, the model is relatively less known, but it has experienced a re-surge of interest in last two decades as a reasonable compromise between realism and simplicity \cite{Zhou97,Taylor95,Taylor03,Taylor09_a,Taylor09_b}. In this model, the polymer is formed by a sequence of consecutive monomers, represented by impenetrable hard-spheres, so that consecutive monomers are tangent to one another, and non-consecutive monomers additionally interact via a square-well interaction. The model can then be reckoned as a variation of the usual freely-jointed-chain \cite{Grosberg94}, with the additional inclusion of a short-range attraction between different parts of the chain and excluded volume interactions. 

In spite of its simplicity, this model displays a surprisingly rich phase behavior, including a coil - globule and a globule - crystal transitions,
that are the strict analog of the gas-liquid and liquid-solid transitions, respectively. Interestingly, even in this case a direct coil - crystal transition is found for sufficiently short range attraction, pushing this analogy with the above direct gas-solid freezing transition even more \cite{Taylor95,Taylor09_a}. 

The SW polymer model can also be easily adapted to mimic the folding of a protein, rather than a polymer. The crucial difference between polymers and proteins
stems from the specificities of each amino acids forming the polypeptide chain that, along with the steric hindrance provided by the side chains, drastically
reduces the number of possible configurations of the folded state \cite{Finkelstein02}. As a result, one obtains a unique native state, rather than a multitude of
local minima having comparable energies. The simplest way to introduce the selectivity defined by different amino acids is given by partitioning the monomers in two
classes, having hydrophobic (H) and polar (P) characters. Under the action of a bad solvent and/or for low temperatures, the H monomers will tend to bury themselves
inside the core of the globule, in order to prevent contact with the solvent (typically water). The HP model has been shown very effective in on-lattice studies 
\cite{Wust08,Seaton09,Wust11}, to describe the folding process, at least at qualitative level. The SW model is also reminiscent of Go-like models 
\cite{Taketomi75,Clementi00,Koga00,Badasyan08} routinely adopted in protein folding studies, where the amino acid specificities are enforced by including the native contact list into the simulation scheme.

One of the main difficulties involved in numerical simulations of polymer chains, stems from the very large computational effort 
necessary to investigate its equilibrium properties for sufficiently long polymers. This is true both using conventional canonical techniques \cite{Allen87,Frenkel02}, 
and more recently developed micro-canonical approaches, such as the Wang-Landau method \cite{Wang01}. Even in the simple SW polymer, while high temperature behavior poses little difficulties, low-temperature/low-energy regions are much more problematic, and yet most interesting. With canonical ensemble simulations the system frequently gets trapped into metastable states at low temperatures, and with the Wang-Landau method the low temperature results strongly depend on the lowest (ground) state energy definition, and an extension of the value of ground energy state to lower values requires increasingly large computational effort. 

It is then of paramount importance to investigate the possibility of using such models for shorter chains and to make a critical assessment on the reliability of the corresponding results. The present paper presents a first step in this direction. More specifically, our aims are
two-fold. First, we will perform a parallel investigation of the SW model for relatively short chains (up to 32 monomers) using both replica exchange
canonical Monte Carlo simulations \cite{Frenkel02} and Wang-Landau micro-canonical technique \cite{Wang01}. A second goal of the paper is to critically
assess the possibility of inferring the full phase diagram, even for these relatively short chains.   

The remaining of the paper is organized as follows. In Section \ref{sec:model}, we will introduce the model and the relevant thermodynamical quantities.
Section \ref{sec:MC} will be devoted to a brief recall of the Monte Carlo simulation techniques, and Section \ref{sec:results} to the obtained results.
The paper will end with some conclusions and perspectives in Section \ref{sec: conclusions}.
\section{Polymer model and thermodynamics}
\label{sec:model}
\begin{figure}[ht]
\includegraphics[width=15cm]{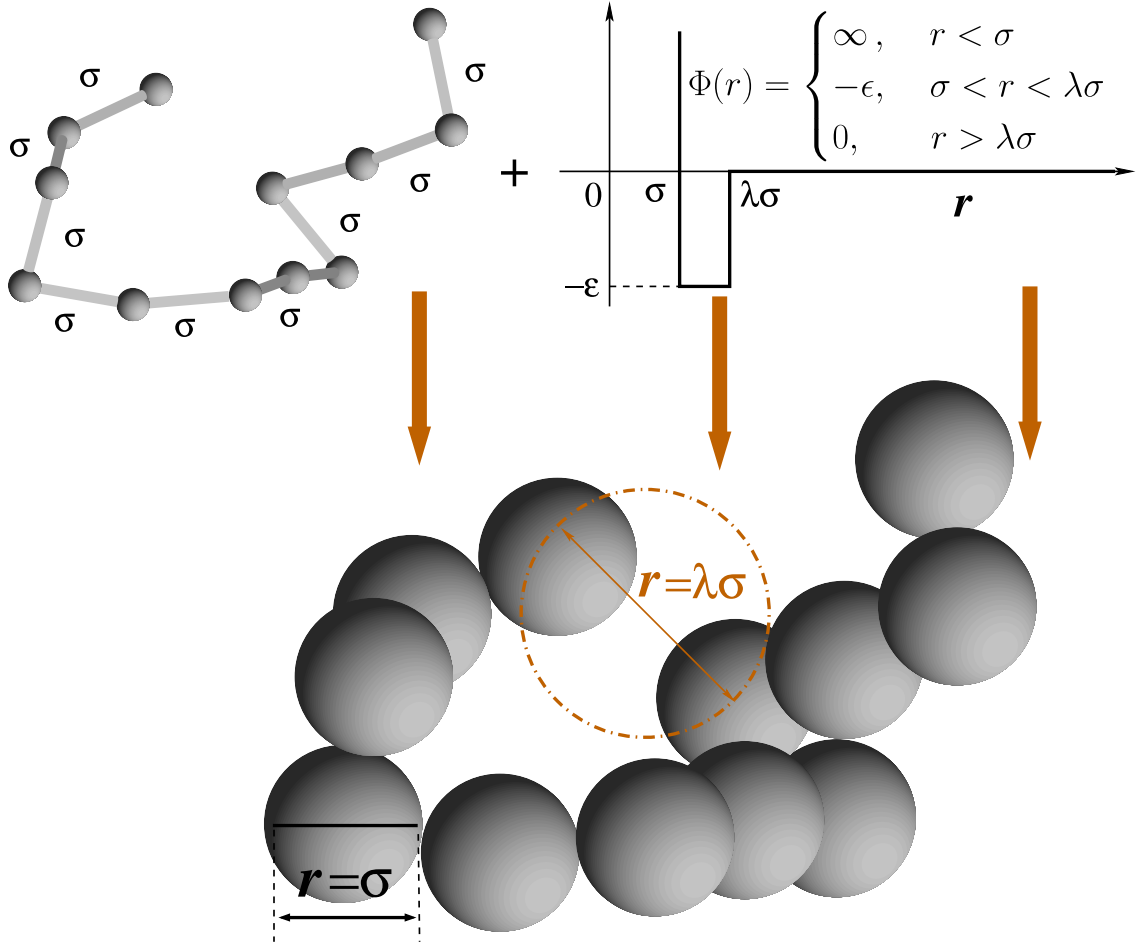}  
\caption{\label{nf1} A homopolymer with square-well potential. The chain can be represented as a set of 
identical connected hard-spheres (diameter $\sigma$) with bond lengths $l=\sigma$. Non-consecutive spheres interact via a square-well potential of range $\lambda \sigma$. Consecutive spheres are tangent one another but cannot inter-penetrate (when $l/\sigma=1$). }
\end{figure}

Following the standard approach \cite{Taylor09_a,Taylor09_b}, we model the system as a flexible homopolymer chain formed by a sequence of $N$ monomers, located at positions
$\{ \mathbf{r}_1,\ldots,\mathbf{r}_N\}$, each having diameter $\sigma$ (see Fig. \ref{nf1}). Consecutive monomers are
connected by a tethering potential keeping the $N-1$ consecutive monomers at fixed bond length $l$. 
Non-consecutive monomers are subject to the action of a square-well 
(SW) potential  
\begin{equation}
\phi(r)=\begin{cases}
+\infty\,,\quad \,r < \sigma&\\
-\epsilon,\quad\, \sigma < r < \lambda \sigma&\\
0, \qquad r > \lambda \sigma&
\end{cases}
\label{sw0}
\end{equation}
\noindent where $r_{ij}=|\mathbf{r}_{ij}|=|\mathbf{r}_j-\mathbf{r}_i|$, and $\lambda-1$ is the well width in units of $\sigma$. Here $\epsilon$ defines the well depth and 
thus sets the energy scale. The model has a discrete spectrum given by $E_n=-\epsilon n$, where $n$ is the number of SW overlaps \cite{Taylor09_b} that, in turn, depends
upon $\lambda$.
In the present paper we will use $l/\sigma=1$ and values of $\lambda$ in the range $[1.03,1.6]$, although the case $l/\sigma >1$ has proven 
to be interesting \cite{Magee08} too.
In micro-canonical approach, a central role is played by the density of state (DOS) $g(E)$ that is related to the micro-canonical entropy 
\begin{eqnarray}
\label{entropy}
S\left(E\right) &=& k_B \ln g\left(E\right),
\end{eqnarray}
($k_B$ is the Boltzmann constant) and hence to the whole thermodynamics. Here additional interesting 
quantities to infer the character of the transition are the inverse micro-canonical temperature \cite{Schnabel11} 
\begin{eqnarray}
\label{beta}
\beta\left(E\right)&\equiv& \left(k_B T\left(E\right)\right)^{-1} = \frac{d S\left(E\right)}{d E}
\end{eqnarray}
and its derivate
\begin{eqnarray}
\label{gamma}
\gamma\left(E\right) &\equiv& \frac{d \beta \left(E\right)}{dE} = \frac{d^2 S\left(E\right)}{d E^2}
\end{eqnarray}

Canonical averages can also be computed using the partition function 
\begin{eqnarray}
Z\left(T\right)&=&\sum_{E} g\left(E\right)e^{-E/\left(k_B T\right)},
\label{Z}
\end{eqnarray}
and the probability function
\begin{eqnarray}
P\left(E,T\right)&=&\frac{1}{Z\left(T\right)}g\left(E\right)e^{-E/\left(k_B T\right)},
\label{Prob}
\end{eqnarray}
to find system in a conformation with the energy $E$ and temperature $T$. Helmholtz and internal energies can then be obtained as
\begin{eqnarray}
F\left(T\right)=-k_B T\ln Z\left(T\right) \\
U\left(T\right)=\left \langle E \right \rangle=\sum_{E} E P\left(E,T\right) 
\label{averages}
\end{eqnarray}
where the average $\langle \ldots \rangle$ is over the probability (\ref{Prob}).

As we shall see, two important probes of the properties of polymers are given by the heat capacity
\begin{eqnarray}
C\left(T\right)&=&\frac{dU\left(T\right)}{dT}=\frac{\left \langle E^2 \right \rangle-\left \langle E\right \rangle^2}{k_B T^2}
\label{Cv}
\end{eqnarray}
and by the the mean squared radius of gyration
\begin{equation}
R_g^2=\frac{1}{N}\sum_{i=1}^N (\mathbf{r}_{i} - \mathbf{r}_{cm})^2, 
\end{equation}
where $\mathbf{r}_{cm} = \frac{1}{N} \sum_{i=1}^N \mathbf{r}_{i}$ 
is the center of mass position.
In practice, one constructs the
$P(R_g^2,E)$ distribution of $R_g$ at a given $E$ to get the micro-canonical average  $\langle \ldots \rangle_E$ over this distribution
\begin{eqnarray}
\left \langle R_g^2 \right \rangle_{E}&=&\sum_{R_g} P\left(R_g^2,E\right) R_g^2 \ ,
\label{RgE}
\end{eqnarray}
and then the canonical average
\begin{equation}
\left \langle R_g^2(T) \right \rangle = 
\sum_{E} \left \langle R_g^2\right \rangle_{E} P(E,T) \ .
\label{RgT}
\end{equation}
It is important to remark that derivation of thermodynamics from the $g(E)$ is quite general, and does not depend on the specific method of simulation. This
is therefore the optimal tool to compare different methods and assess the pros and cons of each of them, that is one of the 
aims of the present work.
\section{Monte Carlo simulations}
\label{sec:MC}
In this section, the equilibrium properties of the above model for homopolymer will be computed using micro-canonical and canonical Monte Carlo simulations,
for chain lengths up to $N=32$. This will allow us to ascertain some specificities of each of them, and to recover some known results at a much lower 
computational effort.
\subsection{Micro-canonical approach: Wang-Landau method}
\label{subsec:micro}
Following general established computational protocols \cite{Allen87,Frenkel02}, and refs.~\cite{Taylor09_a,Taylor09_b} for the specificities related to the
polymers, we use Wang-Landau (WL) method \cite{Wang01} to sample polymer conformations according to micro-canonical distribution,
by generating a sequence of chain conformations $a \to b$, and accepting new configuration $b$ with the micro-canonical acceptance probability
\begin{equation}
P_{acc}(a \rightarrow b)=\min{ \left(1,\frac{w_b g(E_a)}{w_a g(E_b)} \right)},
\label{prob}
\end{equation}
\noindent where $w_a$ and $w_b$ are weight factors ensuring the microscopic reversibility of the moves. 

A sequence of chain conformations is generated using a set of Monte Carlo moves, which are accepted or rejected according to Eq.~\ref{prob}. At a randomly chosen 
site(s) we apply with equal probability pivot (acting on valence and torsional angles, chosen randomly), reptation, crankshaft or backbite (sometimes also referred to as 
end-bridging) moves. For the crankshaft $w_b/w_a=1$ always. Those moves generating conformations through valence angle random sampling must include 
the $w_b/w_a=\sin \theta_b /\sin \theta_a$ ratio in the acceptance probability. This is the case, for instance, of the pivot move acting on valence angle, 
as well as of the reptation move, that includes random generation of valence angles at the ends. The backbite move \cite{Taylor09_a,Taylor09_b} is a non-local move where
one of the chain ends is chosen randomly and all the $\mu$ monomers, lying within a $2\sigma$ range, are counted. If $\mu=0$, the move fails, otherwise one of 
these neighbors is randomly selected, the bond preceding is broken, and a new bond is formed between the chosen neighbor site and end. 
As this bond length is usually larger than $\sigma$, 
an additional shift is often required to satisfy the fixed bond constraint, and  the re-built part of the chain is rotated by a further small randomly chosen angle, 
in order to trigger better sampling. As the last action is nothing else than a pivot move, a full weight factor for the backbite move is 
$w_b/w_a=( \mu \sigma \sin \theta_b)/(r\sin \theta_a)$. A single MC cycle contains at least $N$ move attempts, randomly selected. 
A slightly different version of the backbite move has been successfully applied in Refs.~\cite{Taylor09_a,Taylor09_b,Reith10}.
\subsection{Canonical approach: replica exchange}
\label{subsec:canonical}
The parallel tempering (or replica exchange) technique \cite{Swendsen86,Geyer91} is a powerful method for sampling in systems with rugged energy landscape. It allows the system to rapidly equilibrate and artificially cross energy barriers at low temperatures. Furthermore, the method can be easily implemented on a parallel computer. Parallel tempering technique can be used with both Monte Carlo and Molecular Dynamics simulations, but in this study, we apply this technique to Monte Carlo simulation. The method entails monitoring $M$ canonical simulations in parallel at $M$ different temperatures, 
$T_i$, $i=1,2,\ldots,M$. Each simulation corresponds to a replica, or a copy of the system in thermal equilibrium. In individual Monte Carlo simulation, 
new moves are accepted with standard acceptance probabilities given by the Metropolis method:
\begin{eqnarray}
P_{\Gamma_i \rightarrow \Gamma_i^{'}} &=& \min \left(1,\exp \left(\frac{E_i - E_i^{'}}{k_B T_i} \right) \right) \ ,
\end{eqnarray}
where $E_i$ and $E_i^{'}$ are the energies of the present and the new conformations, $\Gamma_i$ and $\Gamma_i^{'}$, respectively. The replica exchange technique 
allows the replicas at different temperatures to swap with each other without affecting the equilibrium condition at each temperature. 
Specifically, for two replicas, $\Gamma_i$ being at $T_i$ and $\Gamma_j$ being at $T_j$, the swap move leads to a new state, in which $\Gamma_i$ is at $T_j$ 
and $\Gamma_j$ is at $T_i$. The acceptance probability of such a move can be derived based on the condition of detailed balance and is given by:
\begin{eqnarray}
P_{(\Gamma_i,T_i)(\Gamma_j,T_j) \rightarrow (\Gamma_j,T_i)(\Gamma_i,T_j)} &=& \min \left(1,\exp \left[ \left(\frac{1}{k_B T_i} - 
\frac{1}{k_B T_j}\right)(E_i - E_j)\right] \right) \ .
\end{eqnarray}
The choice of replicas to perform an exchange can be arbitrary, but for a pair of temperatures, for which replicas are exchanged, the number of swap move 
trails must be large enough to warrant the statistics. The efficiency of a parallel tempering scheme depends on the number of replicas, the set of temperatures 
to run the simulations, how frequent the swap moves are attempted, and is still a matter of debate. It has been suggested that for the best performance, 
the acceptance rate of swap moves must be about 20\% \cite{Rathore05}.

In our parallel tempering scheme we consider 20 replicas and the temperatures are chosen such that they decrease geometrically: $T_{i+1}=\alpha T_i$, 
where $\alpha=0.8$. The highest reduced temperature is $k_B T_1/\epsilon=10$, at which the polymer is well poised in the swollen phase. We allow replica exchange only 
between neighboring temperatures and for each replica a swap move is attempted every 50 Monte Carlo steps. Standard pivot and crank-shaft move sets are used 
in Monte Carlo simulations. A typical length of the simulations is $10^9$ steps per replica.

Results from parallel tempering simulations are equilibrium data and are convenient to be analyzed using the weighted histogram analysis method 
\cite{Ferrenberg89}. The latter allows one to estimate the density of states as well as to calculate the thermodynamic averages from simulation data 
at various equilibrium conditions in such a way that minimizes the statistical errors where the histograms overlap. 
\section{Results}
\label{sec:results}
\begin{figure}[ht]
\includegraphics[width=10cm]{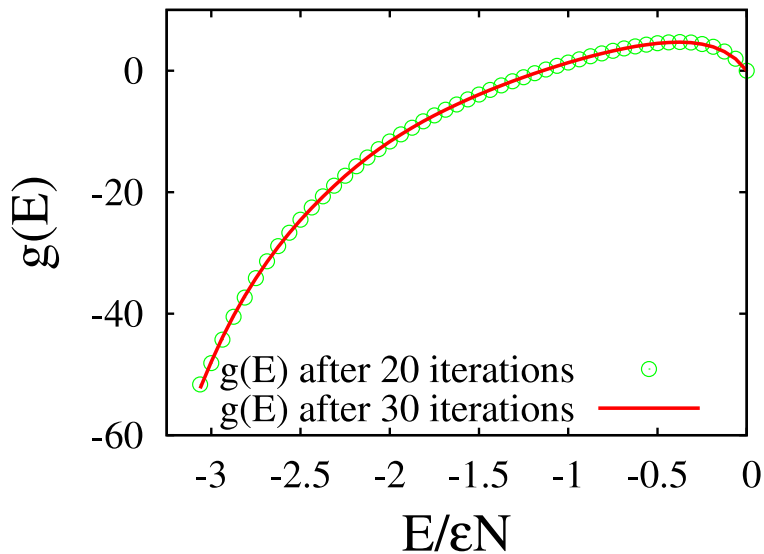} 
\caption{\label{nf2} 
DOS for $N=16$, $\lambda=1.5$ and $E_{min}=-49$ after 20 and 30 iterations. The two results are practically indistinguishable with a normalized root-mean-square-deviation (NRMSD) between the data of $0.002110$.} 
\end{figure}

The aim of the calculation is the computation of the density of states (DOS) $g(E)$ as remarked. In WL method, $g(E)$ is constructed iteratively, 
with smaller scale refinements made at each level of iteration, controlled by the flatness of energy histogram. We typically consider an iteration 
to reach convergence after $26-30$ levels of iteration, corresponding to a multiplicative factor values of $f=1-10^{-8} \div 1-10^{-9}$. 
This choice is neither unique, nor universally accepted, and as this point is crucial for our analysis, it is discussed in some details below.
Seaton and coworkers ~\cite{Seaton09} argued 20 iterations to suffice, while Zhou and Bhatt ~\cite{Zhou05} have additionally shown, that the statistical error of the WL method scales with $f$ as $\sqrt{1-f}$ and thus it is of order of $10^{-3}$ after 20 and $10^{-4}$ after 26 iteration steps. We have explicitly checked this point in our simulations (see Fig.\ref{nf2}),
where the DOS for a chain of $N=16$ is reported for both $20$ and $30$ iterations, with virtually indistinguishable results. As a result, the value 
$26$ was used for most of the subsequent simulations in order to keep the computational effort low. 
However, for longest chain length considered in the present work ($N=32$), we do observe a slight dependence on number of iterations. 
On the other hand, with extending the number of iterations the better quality of data is not guaranteed, since the error saturation plays an increasingly important role. This point has been raised by several groups \cite{Belardinelli07,Swetnam11} with the rule-of-thumb result that an increase of the number of iterations does not necessarily solve the problem since error saturation is an intrinsic feature of flatness-controlled WL simulation. Time-controlled iteration, offered by same authors, appears to improve the situation \cite{Belardinelli07,Swetnam11}, but it gives disappointing results when applied of to a simple hydrophobic-polar (HP) model of protein folding, with some of the low energy states resulting unaccessible for a long time, thus providing unsampled regions of DOS. 
Additional recipes were offered by Swetnam and Allen \cite{Swetnam11}, based on the works of Zhou and Su \cite{Zhou08}, but we find the original flatness-controlled 
algorithm to be more reliable, in the sense that if the algorithm does not converge within reasonable amount of time, as a results of error saturation and/or poor sampling of low-energy states, the correct sampling is clearly affected and the run should be discarded. 

An additional crucial step in WL algorithm hinges in the selection of ground state energy. As this must be defined at the outset, and is known to drastically affect the low-energy behavior of the system \cite{Wust08,Seaton09}, care must be exercised in its selection. At the present time, however, there is no universally accepted procedure for off-lattice Wang-Landau simulations, and in the present paper we will be following the procedure suggested in Refs. \cite{Taylor09_a,Taylor09_b}, that has been reported to be reliable in most of the cases. A preliminary run with no low-energy cutoff is carried out for a sufficient number of MC steps ($10^8N$ in our case). This provides an estimate of the minimal ground state energy. In order to avoid poor sampling and large computational time, this value is increased of few percents (about $2\%$ in our case), and the result is used as the "practical" estimate of the ground state energy (see detailed description in Section II of Ref.~\cite{Taylor09_a}). 

\begin{figure}[ht]
\includegraphics[width=10cm]{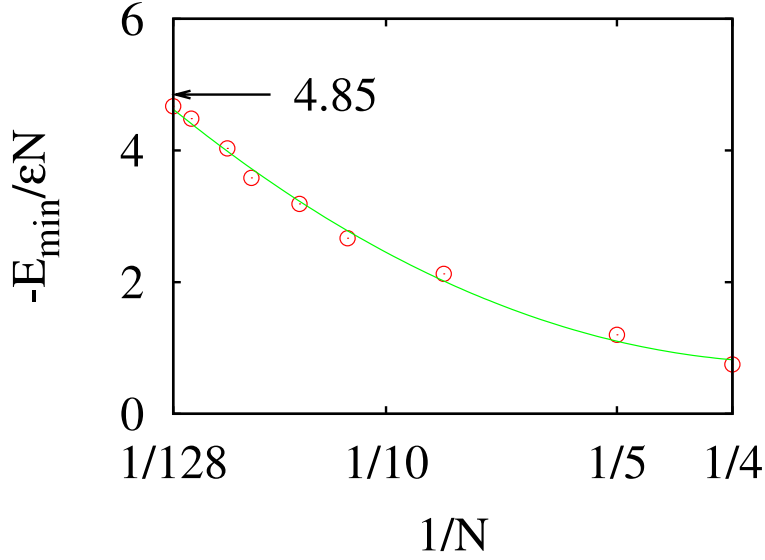}  
\caption{\label{nf3}
Plot of the reduced minimum energy per monomer $-E_{min}/(N \epsilon)$ as a function of $1/N$ for $\lambda=1.5$. 
The extrapolated value to $N \to \infty$ is $-E_{min}/(N \epsilon)=4.85$.} 
\end{figure}

We have explicitly performed this procedure for chains ranging from $N=4$ to $N=128$. This is depicted in Fig.~\ref{nf3}, where the reduced minimum energy per monomer $-E_{min}/(N \epsilon)$ is plotted against $1/N$. There is a clear trend to saturate toward an estimated $-E_{min}/(N\epsilon) \approx 5$ that appears at longer chain lengths. Whether this is related to close packing effects and the symmetry of the ground state conformation, is still 
unclear. Anyway, the lowest energy per unit of chain length appears to saturate to a well defined value at longer chain length.

\begin{figure}[ht]
\includegraphics[width=15cm]{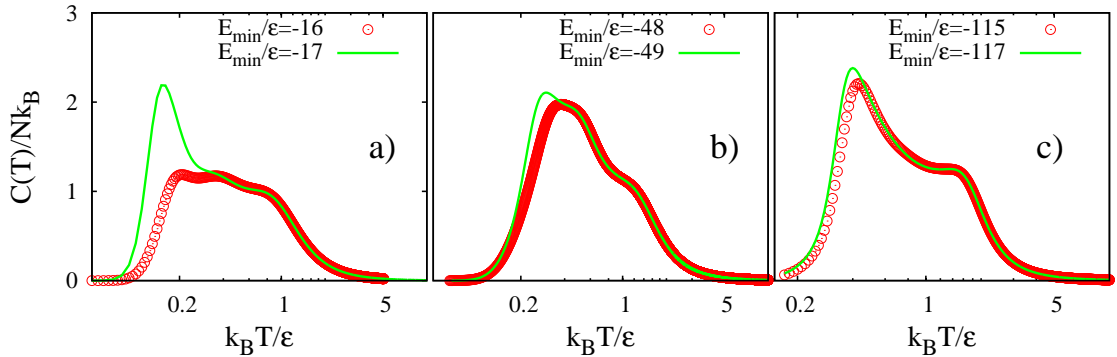} 
\caption{\label{nf4} 
Plot of the reduced heat capacity per monomer $C(T)/(N k_B)$ as a function of $k_BT/\epsilon$ for three different values (from left to right) of polymer length $N=8,16$ and $32$, denoted as a), b) and c), correspondingly, and for different ground state energies. In all cases  $\lambda=1.5$.}
\end{figure}

To illustrate the danger of using an incorrect value to predict the DOS, and quantify its dependence with increasing length of the polymer (that is with increasing $N$),
in Fig.~\ref{nf4} we report the calculation of the heat capacity given by Eq.(\ref{Cv}) for two different choices of $E_{min}/\epsilon$ and different number of 
monomers $N$, as a function of the reduced temperature $k_BT/\epsilon$. Clearly, the low temperatures region is significantly affected by a different choice 
of the ground state, with an error gradually decreasing with $N$. For instance, when $N=8$ cases, the difference between $E_{min}/\epsilon=-16$ and 
$E_{min}/\epsilon=-17$ is larger than with $E_{min}/\epsilon=-115$ and $E_{min}/\epsilon=-117$ when $N=32$. 

As a preliminary step, we have tested our code against exact analytical results valid for small $N$, \cite{Taylor95,Taylor03,Magee08}, and against
previous results using other techniques \cite{Zhou97}. In all cases, we found very good agreement as detailed below.

\begin{figure}[ht]
\includegraphics[width=10cm]{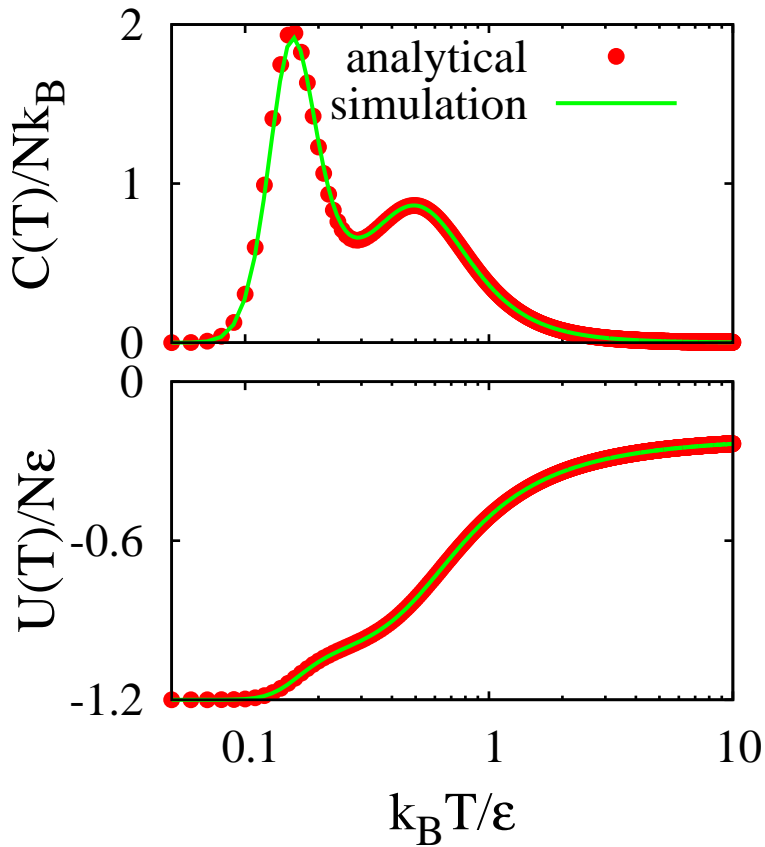} 
\caption{\label{nf5} 
Plot of the reduced heat capacity per monomer $C(T)/(N k_B)$ and of the internal energy per monomer $U(T)/N$ as a function of $k_BT/\epsilon$ from our WL code
and the exact results in the case of $N=5$. Here $\lambda=1.4$, and $E_{min}/\epsilon=-6$.} 
\end{figure}

Taylor \cite{Taylor95,Taylor03}, and later Magee \emph{et al} \cite{Magee08}, have computed the DOS analytically for short chain lengths $N=4,5,6$. Using the MATHEMATICA system Ref.~\cite{math}, we have reproduced the results for tetramers and pentamers at different values of $\lambda$ reported in Tables 1 and 2 of Ref.~ \cite{Magee08}, as well as in Figs. 2,3 and 4 of Ref.~ \cite{Taylor03}. With this being done, we have then compared results from our simulation code for the same $N=4,5$, always finding an excellent agreement. As example of this with $N=5$ (pentamer) and $\lambda=1.5$, is reported in Fig.~\ref{nf5}, both for the reduced heat capacity per monomer $C(T)/(N k_B)$ and for the internal energy per monomer $U(T)/(N \epsilon)$. 

\begin{figure}[ht]
\includegraphics[width=10cm]{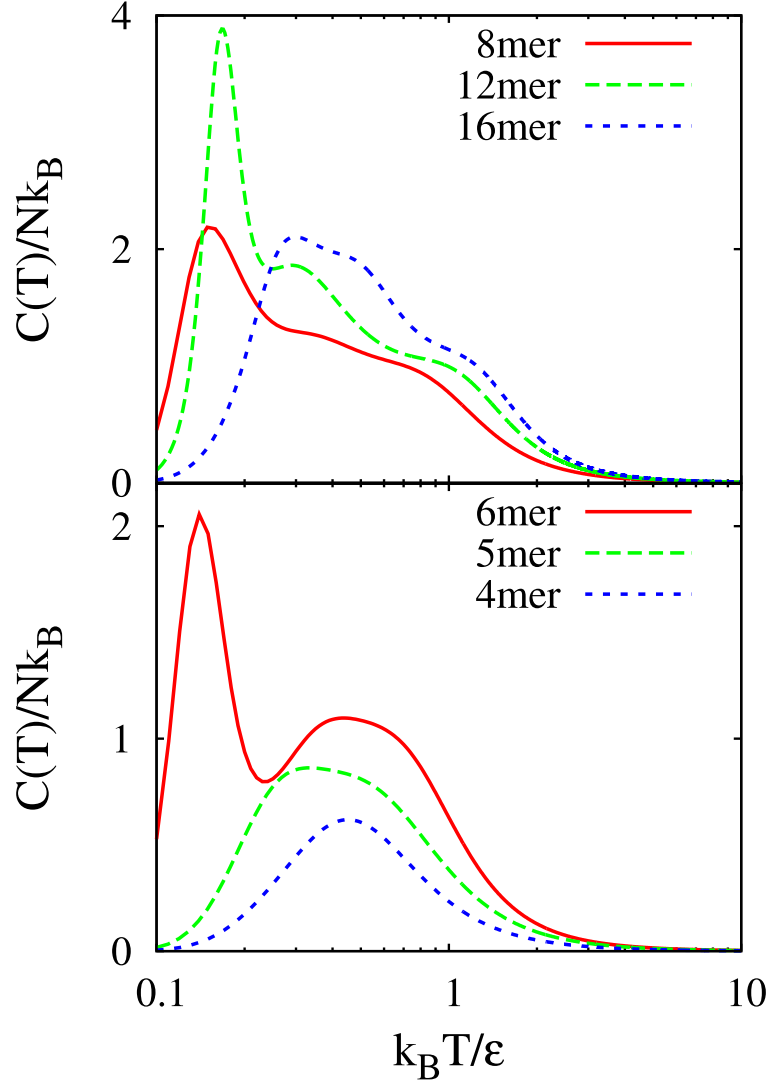} 
\caption{\label{nf6} 
Reduced heat capacity per monomer $C(T)/(N k_B)$ for $N=4,5,6,8,12,16$ as a function of $k_BT/\epsilon$, at $\lambda=1.5$. These findings are in very good agreement with 
Zhou and Karplus (see Fig.3 in Ref. \cite{Zhou97}). } 
\end{figure}

Further support to the correctness of our WL code stems from a comparison with the results by Zhou and Karplus \cite{Zhou97}, who studied the same model using discontinuous molecular dynamics (DMD) and canonical MC simulations. Some test runs for small chains $N=4-16$, are reported in  Fig.~\ref{nf6}, and can be checked against Fig.3 in Ref.~ \cite{Zhou97}. In all cases, a very good agreement is found for the heat capacity per monomer 
$C(T)/(N k_B)$, that is known to be a very sensible probe. 
Note that, in addition to the well-expressed maximum, last three chain lengths ($N=8,12,16$) indicate the presence of two less distinct maxima/plateaux.
It is worth stressing that in the case $N=16$, our calculation is able to probe lower temperature regions than in Ref.\cite{Zhou97}, thus enlightening the appearance of the maximum that results blurred in Ref.\cite{Zhou97}. This is because, results from Ref.\cite{Zhou97} are affected by very large errors at low temperatures (as commonly found in canonical calculations) even for moderate chain lengths. For instance, their case $N=16$ shows the onset of a large error below $k_B T/\epsilon=0.3$, so lower temperatures are in fact cut out (see again their Fig.3). Conversely, in WL calculation, low and high temperatures are equally well sampled and, because of this, we have managed to highlight the appearance of peak in the heat capacity at low temperatures, that is an indicator of a possible structural transition (see Fig. ~\ref{nf6}).
WL approach is therefore extremely useful in this respect. As we have mentioned, Zhou and Bhatt \cite{Zhou05} estimated the convergence error in WL algorithm to scale 
as $\sqrt{1-f}$, so a conservative estimate of the error is of the order $10^{-4}$. 

\begin{figure}[ht]
\includegraphics[width=15cm]{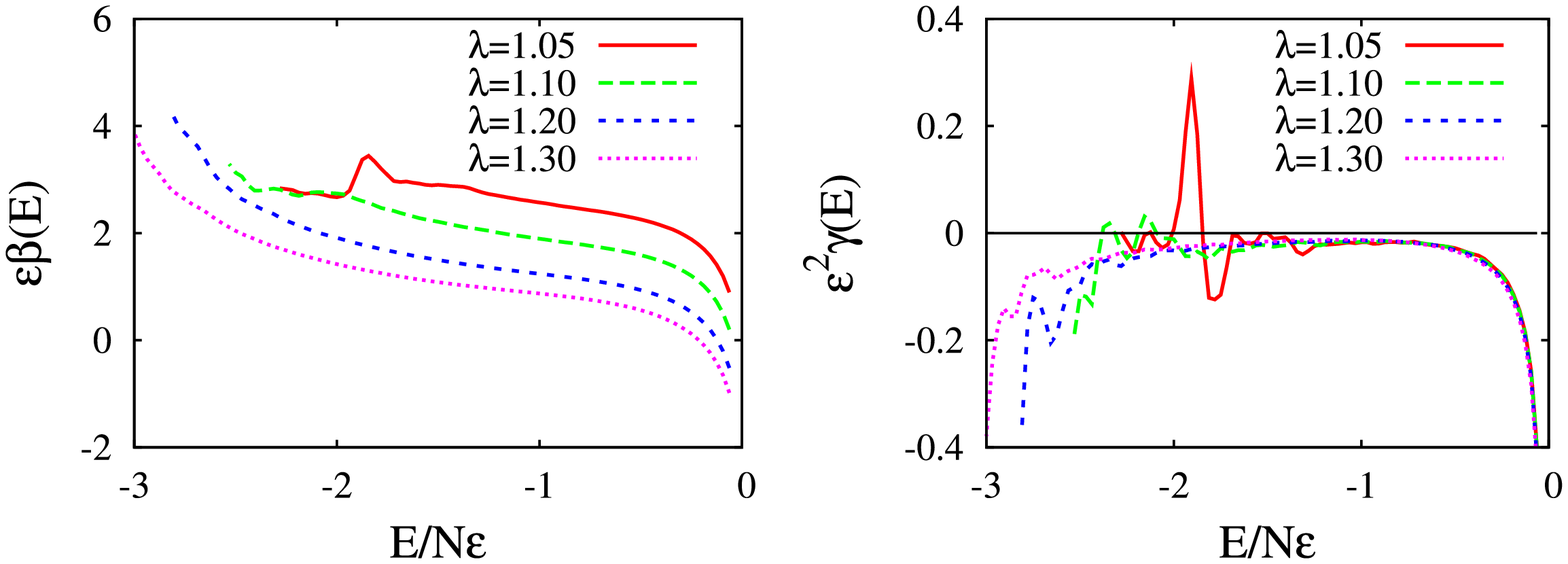} 
\caption{\label{nf5bis} 
Plot of the reduced first ($\epsilon \beta(E)$) and second ($\epsilon^2 \gamma(E)$) derivative of the inverse micro-canonical temperature, as a function of the reduced energy per monomer $E/(N \epsilon)$. Considered values of $\lambda$ are $1.05,1.10,1.20,1.30$.} 
\end{figure}

Additional insights within the micro-canonical approach can be obtained by computing the first $\beta(E)$ and second $\gamma(E)$ derivative of the
inverse micro-canonical temperature as given in Eqs.(\ref{beta}) and (\ref{gamma}), as they are known to be good proxies of structural changes,
with also the possibility to distinguish between first and second order transitions \cite{Schnabel11}.
In particular, the extrema of the second derivative $\gamma(E)$ indicate the corresponding transition energies, with negative and positive values
associated with second and first order transition, respectively.
Our results are depicted in Fig.\ref{nf5bis}, where both these quantities are computed as a function of the reduced energy per monomer $E/(N \epsilon)$.
The clear difference between the $\lambda=1.05$ and the remaining other values of $\lambda$ is a reflection of the existence of a direct
coil-crystal transition, without passing through an intermediate globular state, that was discussed by Taylor \emph{et al} \cite{Taylor09_a},
\cite{Taylor09_b} with much longer chains.

\par
\begin{figure}[ht]
\includegraphics[width=10cm]{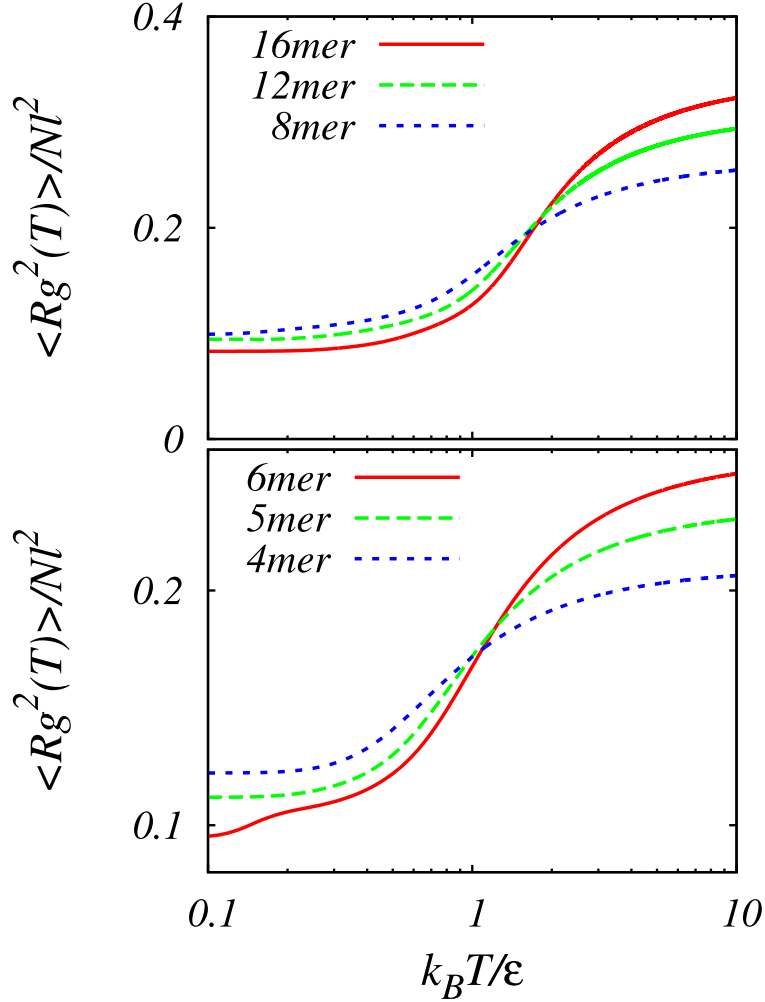} 
\caption{\label{nf7} 
Mean square radius of gyration $\langle R_g^2 \rangle /(N l^2)$ as a function of $k_B T/\epsilon$ and different length chains, 
and for chain lengths $N=4,5,6,8,12,16$ at $\lambda=1.5$. 
One can compare these results against Fig.1 of \cite{Zhou97}.} 
\end{figure}

 In line with other previous results, we note that the radius of gyration is not a good probe for a detailed description of structural changes occurring in the chain at low temperatures, since globule -- crystal transition does not involve any significant change of average size. This can be seen, for instance, in Fig.~\ref{nf7}, where the mean square radius of gyration per monomer $\langle R_g^2 \rangle /(N l^2)$ is plotted as a function of the reduced temperature $k_BT/\epsilon$ for the same chain lengths as above. While it can be clearly seen that each polymer experience a significant shrinks in the region of temperatures between $1.0$ and $3.0$, that is in the same interval where a very weak high-temperature peak/plateau is observed on heat capacity curve (see Fig.~\ref{nf6}), no noticeable changes occur on further cooling down, as opposed to the heat capacity, that shows additional, well pronounced peaks at low temperatures. 
\par As anticipated, one of our main aims was a careful comparison between micro-canonical and canonical approach, in order to assess the strengths and weaknesses 
of each method in the respective domains, and on the reliabilities of results obtained for short chains.
Therefore, we have cross-checked our results with specialized Monte Carlo simulations in the canonical ensemble, using parallel tempering and replica exchange 
improvements for chain lengths up to $N=32$. For chain lengths $N=5,8$ and $12$ the comparison with exact and WL solutions are indistinguishable. 
The main advantage of the canonical method, as compared to the WL counterpart, lays on the fact that we do not have to guess the ground state from the outset, 
as it will naturally emerge as an equilibrium state at sufficiently low temperatures. The drawback for longer chains is, of course, that at low temperatures 
the system becomes more and more compact, and a correct sampling becomes increasingly difficult. 

\begin{figure}[ht]
\includegraphics[width=7cm]{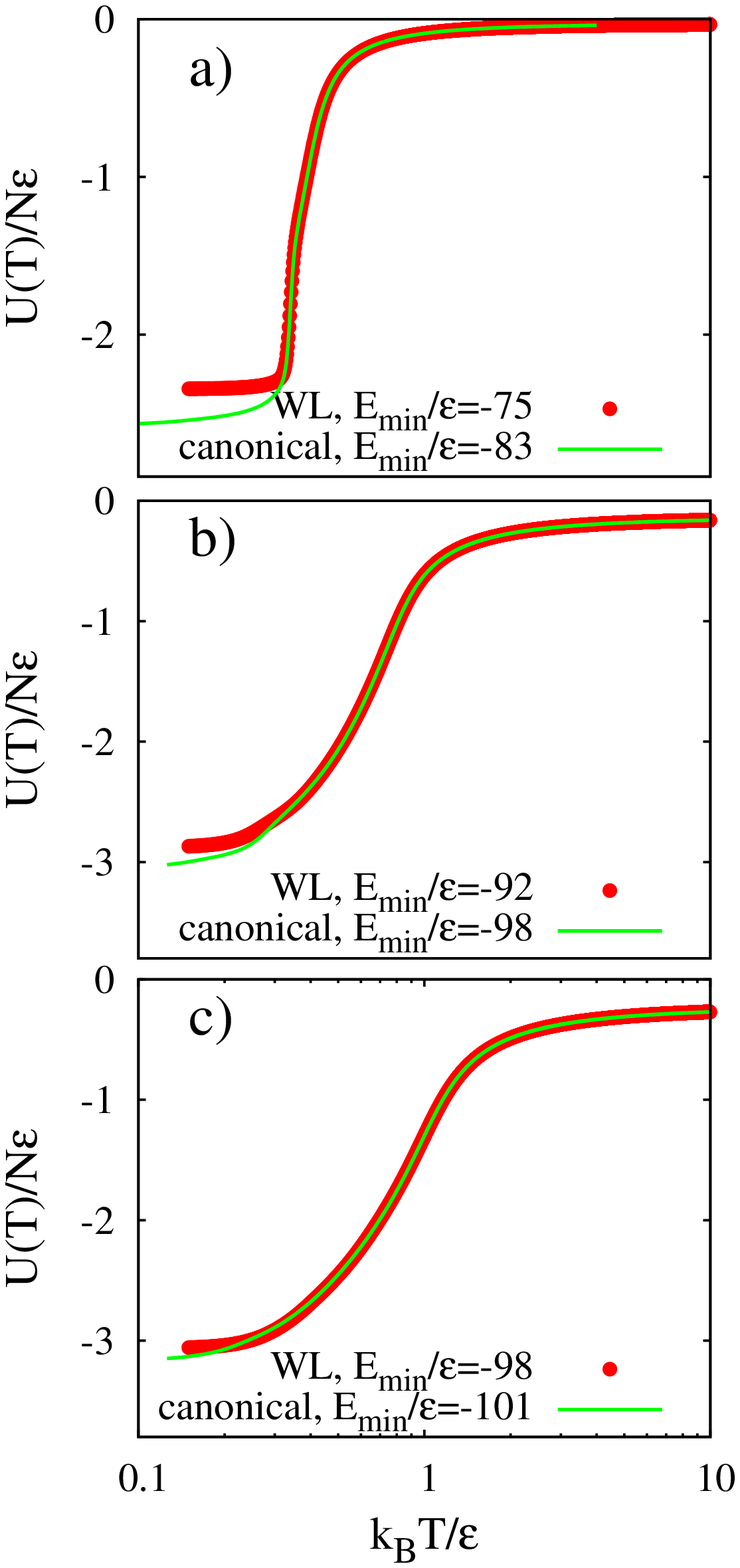} 
\caption{\label{nf8} Plot of of the internal energy per particle $U(T)/(N \epsilon)$, as a function of $k_B T/\epsilon$, for three different values of the parameter $\lambda$: (a) $\lambda=1.05$, (b) $\lambda=1.2$, (c) $\lambda=1.3$. In all cases $N=32$. Both the WL and the canonical results are reported.} 
\end{figure}
\begin{figure}[ht]
\includegraphics[width=10cm]{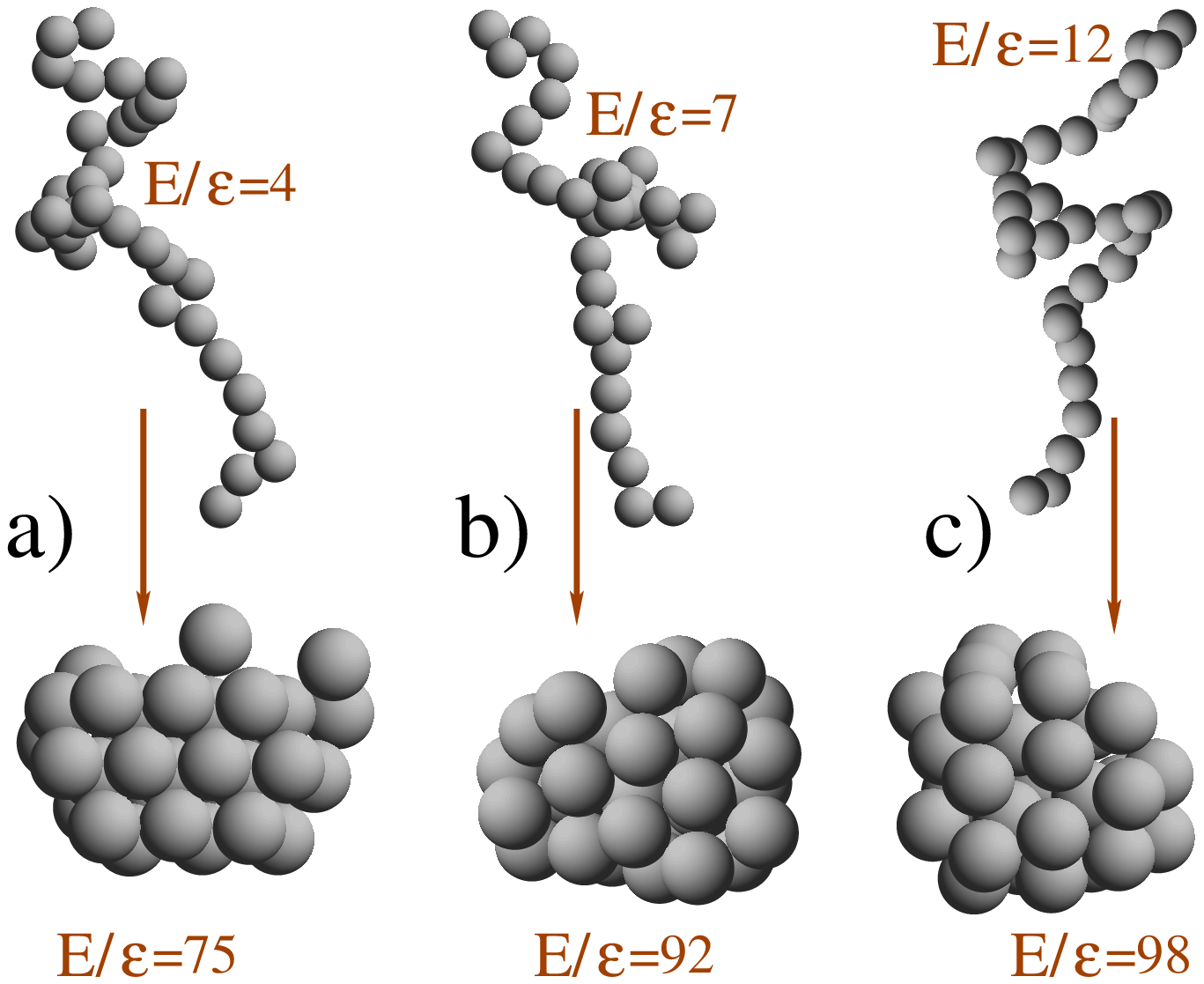} 
\caption{\label{nf8bis} Representative snapshots of coil (top) and globule (bottom) are shown, for three different values of the parameter $\lambda$: (a) $\lambda=1.05$, (b) $\lambda=1.2$, (c) $\lambda=1.3$. In all cases $N=32$. Energies for each of the structures shown on the figure.
} 
\end{figure}
\begin{figure}[ht]
\includegraphics[width=7cm]{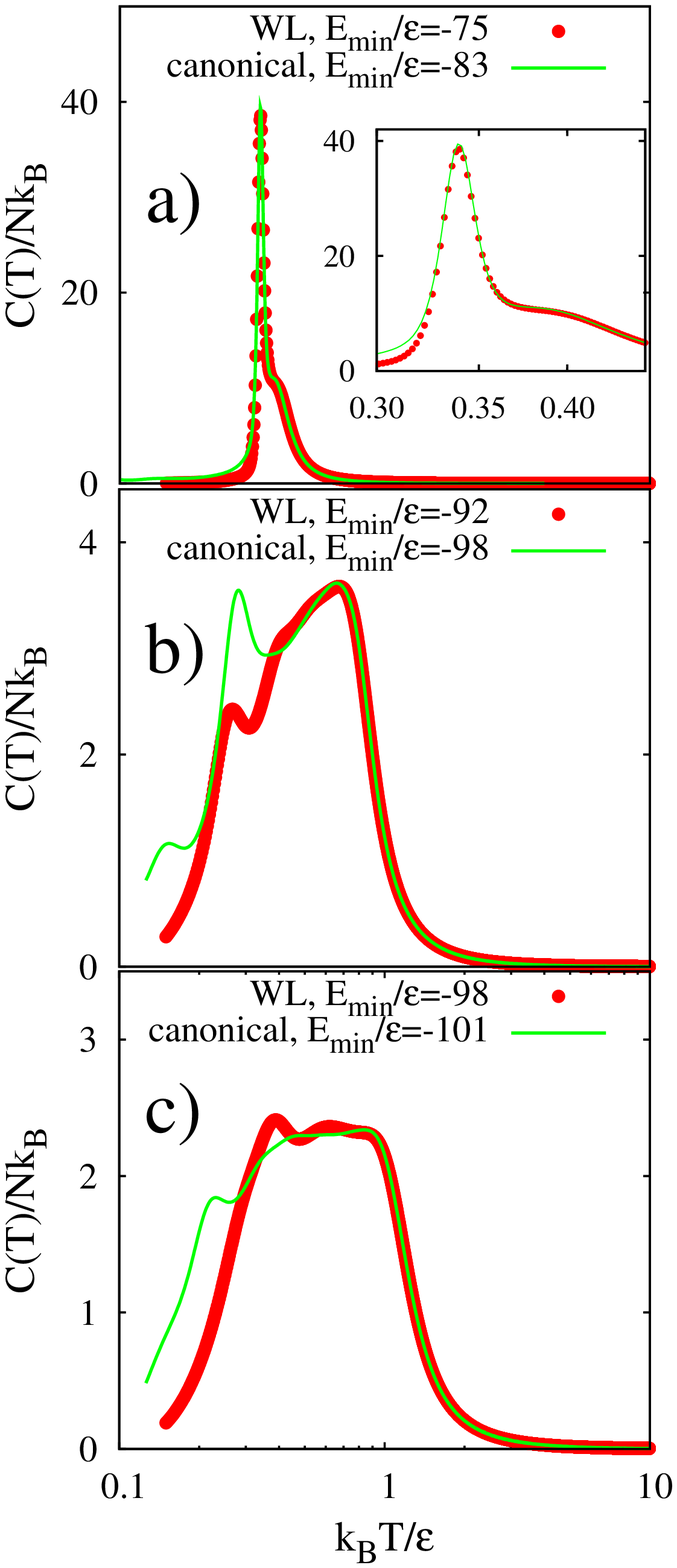} 
\caption{\label{nf9} Plot of the temperature dependence of the reduced heat capacity per particle $C(T)/(N k_B)$, for three different
values of the parameter $\lambda$: (a) $\lambda=1.05$, (b) $\lambda=1.2$, (c) $\lambda=1.3$. In all cases $N=32$. Both WL and canonical
results are reported. This inset in (a) is a blow-up of the most representative part of the figure.} 
\end{figure}

In this parallel calculations, we find a significant difference between the ground state energies as computed from the canonical and the Wang-Landau method, at all considered values of $\lambda$. We further note that for temperatures just above the ground state, the curvature of temperature dependence of the internal energy in low-temperature region is slightly different in the two ensembles, although the absolute values of the energy look quite similar. 
This is reported in  Fig.~\ref{nf8} (b) and (c), with representative snapshots of the initial and final state stemming from the WL calculations depicted in Fig.\ref{nf8bis}. This difference is magnified in the computation of the heat capacity, as shown in Fig.\ref{nf9}, where one can clearly see that for the cases $\lambda=1.2$ (b) and $\lambda=1.3$ (c), peak locations and heights differ. We interpret the first peak (high temperature) in the heat capacity to correspond to the coil-globule transition, and this is supported by the previous results on the mean radius of gyration in Fig.~\ref{nf7}. The additional peaks appearing at lower temperatures are indications of further globule-globule structural changes, not mirrored by the radius of gyration, as remarked. 

It can be easily checked, by matching the energies  $U(T)$ corresponding to the temperature transitions in Fig. \ref{nf8}, that they nicely match the transition energies reported in Fig.\ref{nf5bis} from the micro-canonical approach. 

We note the small discrepancy between the low energy behaviors as obtained from WL and from canonical approaches. This is of course always possible in such
both have advantages and disadvantages that are somewhat complementary one-another. The canonical ensemble calculation is more efficient in predicting the correct absolute value of the ground state energy, as the system is naturally driven toward the absolute minimum by the annealing process, unlike the WL scheme where this is approximately estimated during the initial run. When complemented by replica exchange techniques, allowing a constant swapping of conformation between high and low temperatures, the canonical scheme has proven very reliable in the correct sampling of configuration space at all temperatures. Of course, the sampling becomes increasingly difficult at low temperatures due to the non-swapping moves, as remarked. In WL scheme, on the other hand, sampling is equally achieved at all energies belonging to the chosen interval, including those typically occurring at low temperatures. On the other hand, if the lowest energies are not correctly accounted for, one loses an important contribution from those states in thermodynamical averages, which can be very well more significant at low temperatures. 

Although results obtained with WL and canonical methods slightly differ from one another at low temperatures, they both provide the same physical picture indicating the appearance of two transitions on the phase diagram: coil -- molten globule at higher temperatures and molten globule -- globule at lower temperatures. 
In spite of the much shorter polymer lengths, our results are in qualitative agreement with those obtained in Refs.~\cite{Taylor09_a,Taylor09_b} 
for much longer polymers ($N=128$). As in that case, indeed, at values of $\lambda$ close to unit unity, there is only one, 
coil -- crystal transition, as indicated by the inset in Fig.\ref{nf9} (a).

\begin{figure}[ht]
\includegraphics[width=15cm]{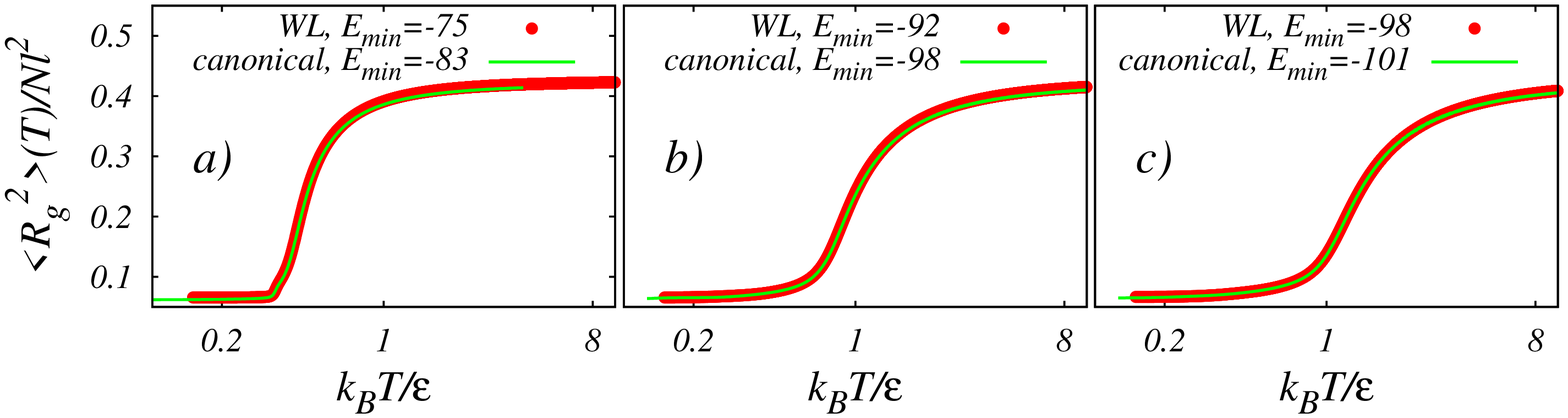} 
\caption{\label{nf10} Plot of the temperature dependence of the radius of gyration per particle $\langle R^2_g \rangle/(N l^2)$, 
for three different
values of the parameter $\lambda$ (left-to-right): (a) $\lambda=1.05$, (b) $\lambda=1.2$, (c) $\lambda=1.3$. In all cases $N=32$. Again, both results from the
WL and the canonical approaches are displayed. 
} 
\end{figure}
We have also contrasted results from WL and canonical approach for the average radius of gyration per monomer 
$\langle R_g^2 \rangle/(N l^2)$ as a function
of temperature. This is reported in Fig.\ref{nf10} for the same parameters as above. Two points are noteworthy. 
First the critical temperature of the coil-globule transition is in agreement with those predicted in Figs.\ref{nf8} and \ref{nf9} in both approaches. 
Second, both results indicate that the radius of gyration does not display any significant change in size beside the first coil-globule transition, as one should expect as an indication of globule -- crystal transition, again in contrast with the previous picture hinging upon 
the behavior of heat capacity. In this respect, we thus confirm that the radius of gyration is not a very good probe of these type of transitions.
However, for the case of $\lambda=1.05$ one can notice a sharp change of the radius of gyration near the transition temperature $T=0.34$, in contrast to 
the gradual changes of $R_g$ in the two other cases. This is a manifestation of the first-order like direct transition from coil 
to compact crystal-like phase observed for this value of $\lambda$. 
\section{Conclusions}
\label{sec: conclusions}
In this paper we have studied the equilibrium statistics of a homopolymer formed by a sequence of tangent identical monomers represented by impenetrable hard spheres.
Non-consecutive spheres, interact via a square-well potential thus driving the collapse of the chain at sufficiently low temperatures.
Both Wang-Landau micro-canonical and replica-exchange canonical calculations were performed for polymers up to $N=32$ monomers. We have then privileged 
cross-checking between different approaches over extensive simulations of very long chains. In this respect, our approach is complementary to those
carried out by Taylor \emph{et al} \cite{Taylor09_a,Taylor09_b}, where much longer chains (up to $N=256$ within a single approach) 
were studied. This comparison allows to uncover the pros and cons of each approach for short chains where, presumably, an exhaustive comparison can be carried out. Our results are in complete agreement with those from Taylor \emph{et al} \cite{Taylor09_a,Taylor09_b}, in such we also observe evidence of a double transition 
coil--globule at higher temperatures, and globule -- crystal transitions at lower temperatures, by working at much shorter polymer lengths, and hence with a significant less computational effort involved, that in the work by Taylor 
\emph{et al} \cite{Taylor09_a,Taylor09_b}, is at the edge of present numerical capability.

Several possible perspectives can be envisaged as a continuation of the present work. By allowing consecutive monomers to inter-penetrate, a local stiffness
can be enforced \cite{Zhou97,Magee08}, thus allowing for other possible transitions, in addition to those reported above, including other morphologies
such as helices and tori. A similar effect can be obtained by using a tubular chain that breaks the spherical symmetry of the present model \cite{Banavar07,Poletto08}.
Both these models differ from the present one, in such they might allow for a coil-helix transition, that cannot be obtained by any spherical symmetric potential. They also require higher computational effort and hence must be restricted to small chains only. 
Work along these lines are underway and will be
reported in a future publication.

\begin{acknowledgments}
This work is supported by NAFOSTED grant 103.01-2010.11. A.B. and R.P acknowledge the support of ARRS through grant P1-0055.
\end{acknowledgments}
\bibliographystyle{apsrev}

\end{document}